\documentclass[showpacs,amsmath,amssymb,aps,prd,preprint,groupedaddress,superscriptaddress]{revtex4-1} 
\usepackage[latin1]{inputenc}
\usepackage{graphicx}
\usepackage{tensor}

\begin{document}
\newcommand{\arctanh}{\textrm{arctanh}\,}


\title{Analyzing the radial geodesics of the Campanelli-Lousto solutions}


\author{J. B. Formiga}
\email[]{jansen.formiga@uespi.br}
\affiliation{Centro de Ciências da Natureza, Universidade Estadual do Piauí, C. Postal 381, 64002-150 Teresina, Piauí, Brazil}

\date{\today}


\date{\today}

\begin{abstract}
When dealing with a spacetime, one usually searches for singularities, black holes, white holes and wormholes due to their importance to the motion of particles. There is a family of solution of the Brans-Dicke vacuum equations that has not been fully studied from this perspective. In this paper, I study some properties of this family and find the complete set of solutions that avoids singularity at the point where the metric diverges or degenerates. The possible changes in the metric signature when passing through this point is analyzed. In addition, I also study the radial geodesics and obtain the solutions of some particular cases.
\end{abstract}


\maketitle

\section{Introduction}
There have been many proposals to modify General Relativity in order to obtain a theory with properties such as unifying all the interactions, being a quantizable theory, providing an understanding of the dark energy and dark matter, etc. One of these proposals is the so-called Brans-Dicke theory (BDT) \cite{PhysRev.124.925}, which was first conceived as a Machian theory but now has also been considered an important theory of gravity for other reasons. As examples of these reasons, we can mention the fact that string theory can be reduced to an effective BDT in the low-energy regime \cite{Faraoni_2011,10.1007/s11214-009-9541-6} and used to deal with problems of modern cosmology \cite{PhysRevD.63.043504,PhysRevD.63.107501,PhysRevLett.62.376}. The BDT is certainly one of the most popular alternative theories of gravity and the study of its solutions is of great importance.

An interesting set of solutions of the Brans-Dicke field equations is a three-parameter family of solutions that are static and spherically symmetric. Sometimes this family is referred to as Campanelli-Lousto solutions. These solutions have been analyzed in Refs.~\cite{doi:10.1142/S0218271893000325,10.1007/s10714-005-0181-1,PhysRevD.86.084031} from the perspective of singularities, geodesics,  black holes, wormholes, and experimental tests in the solar system. When analyzing the singularities, the authors missed a solution that has a coordinate singularity. I fill this gap by finding the complete subset of this family that has this kind of singularity. I also analyze the necessary conditions to avoid any change in the signature of the metric when passing through these singularities. 

Although Campanelli-Lousto solutions have already been studied in detail in Refs.~\cite{doi:10.1142/S0218271893000325,10.1007/s10714-005-0181-1,PhysRevD.86.084031}, including an analysis of their wormholes, as far as I know its geodesics have not been worked out yet. In this paper, I not only analyze some general properties of the radial geodesics but also obtain the exact solution of three particular cases.

This paper is organized as follows. In Sec.~\ref{sa}, the aforementioned family of solutions is presented. Section \ref{sb} is devoted to the analysis of the singularities and the possible changes in the signature of the metric, while in Sec.~\ref{sc} the radial geodesics are studied and the exact solution for some cases are obtained. Some remarks are left to Sec.~\ref{sd}.

\section{Brans-Dicke vacuum solution}\label{sa}
Brans-Dicke theory is described by the metric tensor $\tensor{g}{_\mu_\nu}$ and a scalar field $\phi$ that is neither a geometrical field nor a matter one.  The geometrical part of its action can be written in the form
\begin{equation}
S=\int d^4x\sqrt{-g}[\phi R -\frac{\omega}{\phi}\phi_{,\alpha}\phi^{,\alpha}], \label{05032015a}
\end{equation}
where $R$ is the Ricci scalar. By treating $\tensor{g}{_\alpha_\beta}$ and $\phi$ as independent variables, one obtains the following vacuum field equations
\begin{eqnarray}
\tensor{R}{_\alpha_\beta}=\frac{\omega}{\phi^2}\phi_{,\alpha}\phi_{,\beta}+\frac{\phi_{;\alpha\beta}}{\phi}, \label{05032015b}
\\
\Box\phi=0, \label{05032015c}
\end{eqnarray}
where $\tensor{R}{_\alpha_\beta}$ is the Ricci tensor \footnote{It is defined as $\tensor{R}{_\alpha_\beta}=\tensor{R}{^\lambda_\alpha_\lambda_\beta}=\tensor{\Gamma}{^\lambda_\alpha_\beta_,_\lambda}+\ldots$.}, the semicolon stands for covariant derivative, and $\Box$ is the d'Alembertian.

It is well known that Eqs. (\ref{05032015b}) and (\ref{05032015c}) admit the family of solutions (see, e.g., Refs. \cite{doi:10.1142/S0218271893000325,10.1007/s10714-005-0181-1})
\begin{eqnarray}
ds^2=A^{m+1}dt^2-A^{n-1}dr^2-r^2A^nd\Omega^2, \label{05032015d}
\\
A=1-r_0/r, \label{05032015e}
\\
\phi(r)=\phi_0 A^{-(m+n)/2}, \label{05032015f}
\\
\omega=-2(m^2+n^2+nm+m-n)/(m+n)^2,
\end{eqnarray}
where in Eq. (\ref{05032015d}) the symbol $d\Omega^2$ represents the metric on a unit $2$-sphere.

This solution reduces to the Schwarzschild case for $n=m=0$. For $m=-n$ and $(m+1)r_0=2M$, with $M$ being the mass of a spherical body, the metric (\ref{05032015d}) takes the same form as that of Fisher\footnote{The solution (\ref{05032015d})-(\ref{05032015f}) is not exactly the same as that of Fisher because the scalar fields are different. In the case of Eq.~(\ref{05032015f}), the scalar field becomes trivial for $m=-n$.} (see, e.g., Ref. \cite{PhysRevD.83.087502}). For the former case, one can perform a maximal extension and use the range $r\in (0,\infty)$. But for the latter, in general, one takes $(r_0,\infty)$. It seems interesting, therefore, to analyze the range of $r$ for arbitrary values of $m$ and $n$. This will be done at the end of Sec.~\ref{sb}.

\subsection{Wormholes}
From Eq.~(\ref{05032015d}), we see that the relation between the areal radius $R$ and the coordinate $r$ is
\begin{equation}
R=rA^{n/2}, \label{09082015a}
\end{equation}
which, for even values of $n$ ($n=2l$, say), can be rewritten as $r\left| A^l\right|$. Differentiating  Eq.~(\ref{09082015a}), we obtain
\begin{equation}
dr=A^{1-n/2}A_m^{-1}dR, \label{09082015b}
\end{equation}
where $A_m=1-r_m/r$ and $r_m=(1-n/2)r_0$. Notice that $r_m$ is a minimum of $R(r)$ in the interval $[r_0,\infty)$ for $n\leq 0$.

Since Fisher solution possesses wormholes \cite{doi:10.1142/S0218271814500862}, it is natural that the spacetime (\ref{05032015d}) presents wormhole too. To find the values of $m$ and $n$ for which this happens, let us use the definition of a traversable wormhole throat given by Hochberg and Visser in Ref.~\cite{PhysRevD.56.4745}. They define this kind of throat as being a two-dimensional hypersurface of minimal area taken in one of the constant-time spatial slices. This is equivalent to the vanishing of the trace of the extrinsic curvature ($tr(K)$, say) of this hypersurface plus the condition that its derivative with respect to the normal coordinate $n$ be negative. By following the same procedure as that of Ref.~\cite{doi:10.1142/S0218271814500862}, one can write $tr(k)$ and $\partial tr(K)/\partial n$ for a metric of the type
\begin{equation}
ds^2=e^{2\Phi(R)}dt^2-dR^2/\left[1-b(R)/R \right]-R^2d\Omega^2 \label{16082015a}
\end{equation}
as
\begin{equation}
tr(K)=\mp 2\frac{\sqrt{1-b/R}}{R}, \label{16082015b}
\end{equation}
\begin{equation}
\frac{\partial tr(K)}{\partial n_{\pm}}=\frac{2}{R^2} \left[\frac{R}{2}\frac{d}{dR}bR^{-1}+1-b/R \right], \label{16082015c}
\end{equation}
respectively. The upper sign in $\pm$ or $\mp$ is for the region $r>r_m$, while the lower one is for $r_0\leq r<r_m$.

A straightforward calculation shows that, for the metric (\ref{05032015d}), we have $b/R=1-A_m^2/A$. Finally, by using this expression into Eqs.~(\ref{16082015b})-(\ref{16082015c}), we obtain
\begin{equation}
tr(K)=-2\frac{A_m}{\sqrt{A}R} \label{16082015d}
\end{equation}
and
\begin{equation}
\frac{\partial tr(K)}{\partial n_{\pm}}\Biggl|_{r_m}=-\frac{2}{r_mR_m}\left(1-\frac{r_0}{r_m} \right)^{-n/2}<0, \label{16082015e}
\end{equation}
where I have used $\mp|A_m|=-A_m$ in the calculation of $tr(K)$. Since Eq.~(\ref{16082015d}) vanishes at $r_m$ and the inequality in (\ref{16082015e}) holds for $r_m>r_0$, then we have wormholes for $n<0$ (recall that $r_m=(1-n/2)r_0$).

\section{Singularities}\label{sb}
In searching for singularities, let us calculate the scalar invariants. For the metric (\ref{05032015d}) we have
\begin{equation}
I_1=R=\frac{r_0^2}{2}\frac{1}{r^4} \frac{f_1(m,n)}{A^{n+1}}, \label{1032015a}
\end{equation}
\begin{equation}
I_2=\tensor{R}{^\mu ^\nu}\tensor{R}{_\mu _\nu}=\frac{r_0^2}{8r^8}\frac{f_2(r,m,n)}{A^{2(n+1)}}, \label{1032015b}
\end{equation}
\begin{equation}
I_3=\tensor{R}{^\mu ^\nu ^\alpha ^\beta}\tensor{R}{_\mu_\nu_\alpha_\beta}=\frac{r_0^2}{4r^8}\frac{f_3(r,m,n)}{A^{2(n+1)}}, \label{1032015c}
\end{equation}
where $\tensor{R}{^\lambda_\mu_\beta_\nu}$, $\tensor{R}{_\mu_\nu}$, $R$ are the Riemann tensor, the Ricci tensor, and the curvature scalar, respectively.  The functions $f_1$, $f_2$, and $f_3$ are such that they take the following form at $r_0$:
\begin{eqnarray}
f_1=m^2+n^2+nm+m-n, \label{14032015f}\\
f_2(r_0)=r_0^2( -2nm^2-2nm+4mn^2+m^2+5n^2+2m^2n^2+2m^3+2n^3m+m^4+n^4), \label{14032015g}\\
f_3(r_0)=r_0^2(m^4+6mn^2-4nm^2+3m^2n^2-2nm+n^4+m^2+5n^2+2m^3-2m^3n). \label{14032015h}
\end{eqnarray}
To study the singularities, we need to known all the combinations of $m$ and $n$ that make $f_1$, $f_2$, and $f_3$ vanish simultaneously. There is at least one known value in the literature \cite{doi:10.1142/S0218271893000325}, namely, $m=n=0$ (the Schwarzschild case). Nonetheless, there exists another combination that as far as I know had not been found yet. This combination is $n=0$ and $m=-1$. To prove that these two combinations are the only possible ones, we can take the roots of $f_1$ and substitute them into $f_2$ to find the common roots. After that,  we use the common roots of $f_1$ and $f_2$ into $f_3$ to discovery which ones are also roots of $f_3$. Finally, we take the limit of Eqs.~(\ref{1032015a})-(\ref{1032015c}) as $r$ goes to $r_0$ and check that the results are finite. 

It is easy to see that the roots of $f_1$ are
\begin{equation}
m_{\pm}=-\frac{n}{2}-\frac{1}{2}\pm \frac{1}{2}\sqrt {-3n^2+6n+1}. \label{14032015i}
\end{equation}
Using them in Eq. (\ref{14032015g}), one finds that
\begin{equation}
f_2(r_0,\textrm{roots of\ }f_1)/r_0^2=-\frac{3}{2}n^4+\frac{7}{2}n^2+2n^3 \pm\frac{(1+n)n^2}{2}\sqrt {-3n^2+6n+1}, \label{14032015j}
\end{equation}
The roots of $f_2=0$ for the plus sign are $n=0$ ($m=0$) and $n=-1$ ($m=$ complex), while for the minus sign one has $n=0$ ($m=-1$), $n=-1$ ($m$ complex), and $n=2$ ($m=-2$). It is straightforward to verify that the combination $(m,n)=(-2,2)$ is not a root of $f_3(r_0,m,n)=0$, but the other two real combinations are. In short, we have
\begin{equation}
(m,n)= (0,0),\ (-1,0). \label{14032015l}
\end{equation}
Since we already know that there is no singularity for the case $(0,0)$, let us focus on $(-1,0)$. Evaluating the invariants (\ref{1032015a})-(\ref{1032015c}) for this case, we obtain
\begin{equation}
I_1=0,\quad I_2=\frac{3}{2}\frac{r_0^2}{r^6},\quad I_3=6\frac{r_0^2}{r^6}, \label{27042015a}
\end{equation}
which are not singular at $r_0$.

For $m$ and $n$ different from the values (\ref{14032015l}), we can determine the existence of a singularity at $r_0$  by the exponent of $A$ in Eqs. (\ref{1032015a})-(\ref{1032015c}).  In this case, it is clear that there is a physical singularity for $n>-1$, as pointed out in Ref. \cite{doi:10.1142/S0218271893000325}. Therefore, there is no physical singularity at $r_0$ for the cases (\ref{14032015l}) and $n\leq -1$.

To see the topology of $r=r_0$, we can evaluate the area of the surface $r$ and $t$ constant. This gives
\begin{equation}
\textrm{area}=\int_{0}^{2\pi}\int_{0}^{\pi} \sqrt{g_{22}g_{33}}d\theta  d\varphi=4\pi r^2(1-r_0/r)^n, \label{21042014g}
\end{equation}
which leads us to the conclusions:
\begin{eqnarray}
\textrm{null area and topology of a point for}\quad n>0,   \label{21042014h}
\\
\textrm{the finite area $4\pi r_0^2$ for}\quad n=0,   \label{21042014i}
\\
\textrm{infinite area for}\quad n<0.   \label{21042014j}
\end{eqnarray}

If we are to extend the range of $r$ to values smaller than $r_0$, we must avoid the cases where $r_0$ is a physical singularity and also avoid changes in the signature of the metric. It is clear that we cannot extend the range of $r$ if either $m$ or $n$ is an irrational number. There are other values of $m$ and $n$ that can also make the metric complex in the region $r<r_0$ or changes its signature. Generally speaking, we have the following:
\begin{enumerate}
\item The metric becomes complex when either $m$ or $n$ is an irreducible fraction of the type odd/even.
\item Since the term $A^n$ in the angular part of the metric cannot change its sign, $n$ must be either an even integer or an irreducible fraction of the type even/odd. In both cases, the term $A^{n-1}$ changes its sign when passing through $r_0$, which means that $A^{m+1}$ must do the same to prevent change in the signature of the metric. Therefore, $m$ can be either an even integer or even/odd.
\end{enumerate}

It is clear that an analytic extension of the case $m=-1$ and $n=0$ cannot be achieved without changing the signature.

\section{Radial geodesic}\label{sc}
Let us now analyze the behavior of the gravitational field through the radial geodesics. It is well known that if the spacetime admits a Killing vector $k$, then
\begin{equation}
\tensor{k}{_\mu}\tensor{u}{^\mu}=C, \label{14032015a}
\end{equation}
where $C$ is a constant, and $\tensor{u}{^\mu}$ is the $4$-velocity of a test particle that follows a geodesic. One can easily show that $\tensor{k}{^\mu}=\delta^{\mu}_0$ is a timelike Killing vector of the spacetime (\ref{05032015d}). In this case we take $E=C$, where $E$ is the energy of the test particle. Therefore, Eq. (\ref{14032015a})  becomes
\begin{equation}
A^{m+1}\dot{t}=E, \label{14032015b}
\end{equation}
where the overdot means $d/d\tau$ and $\tau$ is the proper time. For material particles, we must also have
\begin{equation}
A^{m+1}\dot{t}^2-A^{n-1}\dot{r}^2=1. \label{14032015c}
\end{equation}
From these two equations, we obtain
\begin{equation}
\Delta\tau=\pm \int_{r_i}^{r}\frac{dr}{\sqrt{E^2A^{-(m+n)} -A^{1-n} }}=\pm r_0\int_{A_i}^{A}\frac{dA}{(1-A)^2\sqrt{E^2A^{-(m+n)} -A^{1-n} }}, \label{14032015d}
\end{equation}
and
\begin{equation}
E^2=A^{m+1}+A^{m+n}\dot{r}^2. \label{14032015e}
\end{equation}

 Using Eq.~(\ref{09082015b}) into (\ref{14032015e}), we get
\begin{equation}
E^2=A^{m+1}+A^{m+2}A_m^{-2}\dot{R}^2. \label{09082015c}
\end{equation}

From Eq.~(\ref{09082015c}) we can infer whether the particle is attracted or repelled from $r_0$. Suppose we release the test particle from rest at $r_i$ with $r_i\neq r_0,r_m$. In this case, Eq.~(\ref{09082015c}) yields $E^2=A_i^{m+1}=(1-r_0/r_i)^{m+1}$ and we can write
\begin{equation}
A_i^{m+1}-A^{m+1}=A^{m+2}A_m^{-2}\dot{R}^2. \label{22032015a}
\end{equation}
For $r>r_0$ we have $A>0$ \footnote{For the case $r<r_0$ and $m$ either an even integer or an irreducible fraction of the type even/odd, the situation is a little bit more involved because setting $\dot{R}_i=0$ no longer means releasing the particle from rest at $r_i$ ($r$ is a time coordinate in this case). I shall not pursue this issue here.}, which means that the right-hand side of Eq.~(\ref{22032015a}) is positive. Since $A$ grows as $r$ increases, it is clear that $r_i\geq r$ for $m>-1$ and $r_i\leq r$ for $m<-1$. In the former case, the particle will be attracted to $r_0$, while in the latter it will be repelled from $r_0$.  

With respect to the case $m=-1$, we have
\begin{equation}
E^2=1+AA_m^{-2}\dot{R}^2. \label{30042015a}
\end{equation}
If $\dot{R}$ is zero at some $r_i$ different from $r_m$, then $E^2=1$, which implies $A\dot{R}^2=0$ for all $r$. Since this last expression must hold for all $r$,  the particle must remain at rest in the initial position. To see what happens when $\dot{R}=0$ at $r_m$, one has to solve (\ref{30042015a}).

By deriving Eq.~(\ref{30042015a}) with respect to $\tau$, one can show that
\begin{equation}
\ddot{R}=(E^2-1)\frac{r_0}{2r^2}A^{-n/2-1}\left[ 1-n-(2-n)\frac{r_0}{2r}\right]. \label{30042015b}
\end{equation}
This expression changes its sign at 
\begin{equation}
r_c=\left(\frac{2-n}{1-n}\right)\frac{r_0}{2}. \label{15082015a}
\end{equation}
It is clear in Eq.~(\ref{30042015b}) that there is no change of sign for either $n=1$ or $n=0$. On the other hand, from Eq.~(\ref{15082015a}) and the fact that $r\geq r_0$, we see that there is a sign change for $0< n <1$ \footnote{Notice that the expression $1-n-(2-n)\frac{r_0}{2r}$ can be recast as $(1-n)(1-r_c/r)$, which clearly changes its sign. }. In this case, the gravitational field is attractive for $r<r_c$ and repulsive for $r>r_c$. However, when we have $n\leq 0$  it is repulsive for all possible values of $r$, while for $n\geq 1$ it is attractive. The non-singular case $m=-1$ and $n=0$ will be analyzed in more detail in Sec.~\ref{s30042015}.

It is worth mentioning that once the particle is in motion, a repulsive force may appear even in the cases where the gravitation field is attractive. An example of such a case is given in Sec.~\ref{s22032015a}. 

\subsection{The case of $m=0$ and $n=-2$}\label{s22032015a}
The substitution of $m=0$ and $n=-2$ into Eqs.~(\ref{14032015e}) and (\ref{14032015d}) leads to
\begin{equation}
E^2=A+A^{-2}\dot{r}^2 \label{21032015a}
\end{equation}
and 
\begin{equation}
\Delta\tau=\pm r_0\int_{A_i}^{A}\frac{dA}{(1-A)^2\sqrt{E^2A^2 -A^3 }}, \label{20032015a}
\end{equation}
which yields
\begin{eqnarray}
\tau=\pm \frac{r_0A}{|A|}\Biggl\{ \frac{1}{E}\ln\left|\frac{A}{(E+\sqrt{E^2-A})^2} \right|+\frac{E^2-3/2}{(E^2-1)^{3/2}}\ln\left[ \frac{(\sqrt{E^2-1}+\sqrt{E^2-A})^2}{1-A}\right]
\nonumber\\
+\frac{\sqrt{E^2-A}}{(1-A)(E^2-1)}\Biggr\} +\textrm{constant}\label{20032015b}
\end{eqnarray}
for $E>1$,
\begin{eqnarray}
\tau=\pm \frac{r_0A}{|A|}\Biggl\{ \frac{1}{E}\ln\left|\frac{A}{(E+\sqrt{E^2-A})^2} \right|-\frac{3-2E^2}{(1-E^2)^{3/2}}\arctan\left(\sqrt{\frac{E^2-A}{1-E^2}} \right)
\nonumber\\
-\frac{\sqrt{E^2-A}}{(1-A)(1-E^2)}\Biggr\} +\textrm{constant}\label{20032015c}
\end{eqnarray}
for $E<1$, and
\begin{equation}
\tau=\pm\frac{r_0A}{|A|}\left\{ \frac{8/3-2A}{(1-A)^{3/2}} -\ln\left[ \frac{(\sqrt{1-A}+1)^2}{|A|}\right] \right\}+\textrm{constant}\label{20032015d}
\end{equation}
for $E=1$.

From these three expressions, one sees that $\tau$ diverges at $r_0$ ($A \to 0$). This means that the time for the test particle to achieve $r_0$ is infinity, whatever the initial conditions are. It is worthwhile to note that the time spent by light to reach $r_0$ as measured by a static observer for $n\leq m$ is also infinite \cite{doi:10.1142/S0218271893000325}. Nonetheless, the interesting point here is not $r_0$ but rather $r_m$. Since $n$ is negative, we are dealing with a wormhole whose throat is at this point.

\begin{figure}[h]
\begin{minipage}[t]{0.45\linewidth}
\includegraphics[scale=0.25]{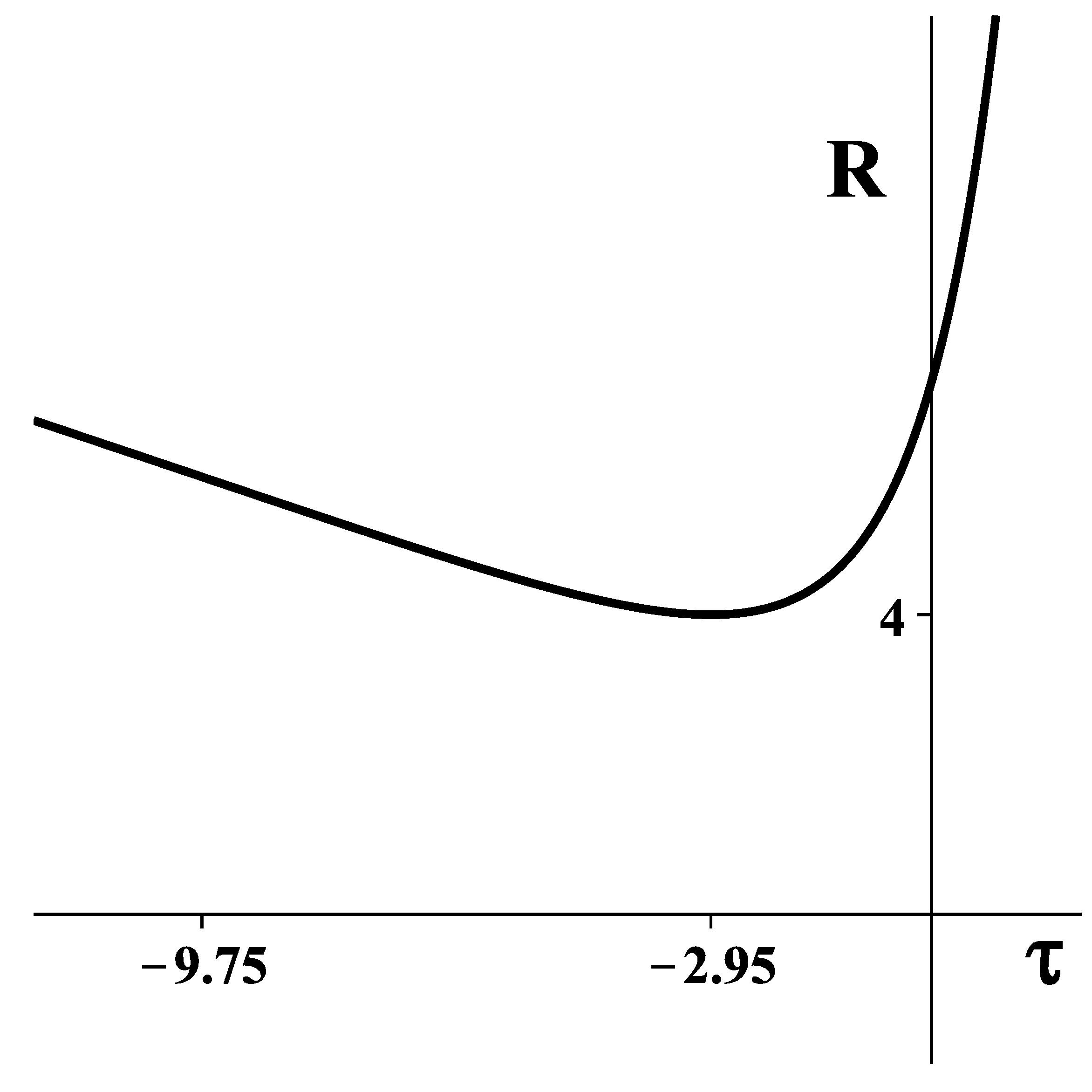}
\caption{\scriptsize  In this plot, the solid curve shows the qualitative  behavior of the curve $(\tau(r),R(r))$  for the case $E=1$ and $\dot{R}=0$ at $r \to \infty$; the axes are in units of $r_0$. The minimum value of $R$ is $4r_0$ and the left side of the minimum in this figure corresponds to $r \in (\infty,2r_0)$, while the right one is due to $(2r_0,r_0)$ (notice that $r_m=2r_0$). As it is clear, the particle is attracted to $r_0$,  nevertheless, a repulsive force appears at $(-9.75r_0, 5.84r_0)$ and the concavity of the curve changes. Notice that the behavior of $R$ is that of a wormhole with a throat at $2r_0$. In this figure and in all the others, the arbitrary constants in the geodesics have been set to zero.}
\label{f21032015a}
\end{minipage}
\hfill
\begin{minipage}[t]{0.45\linewidth}
\includegraphics[scale=0.25]{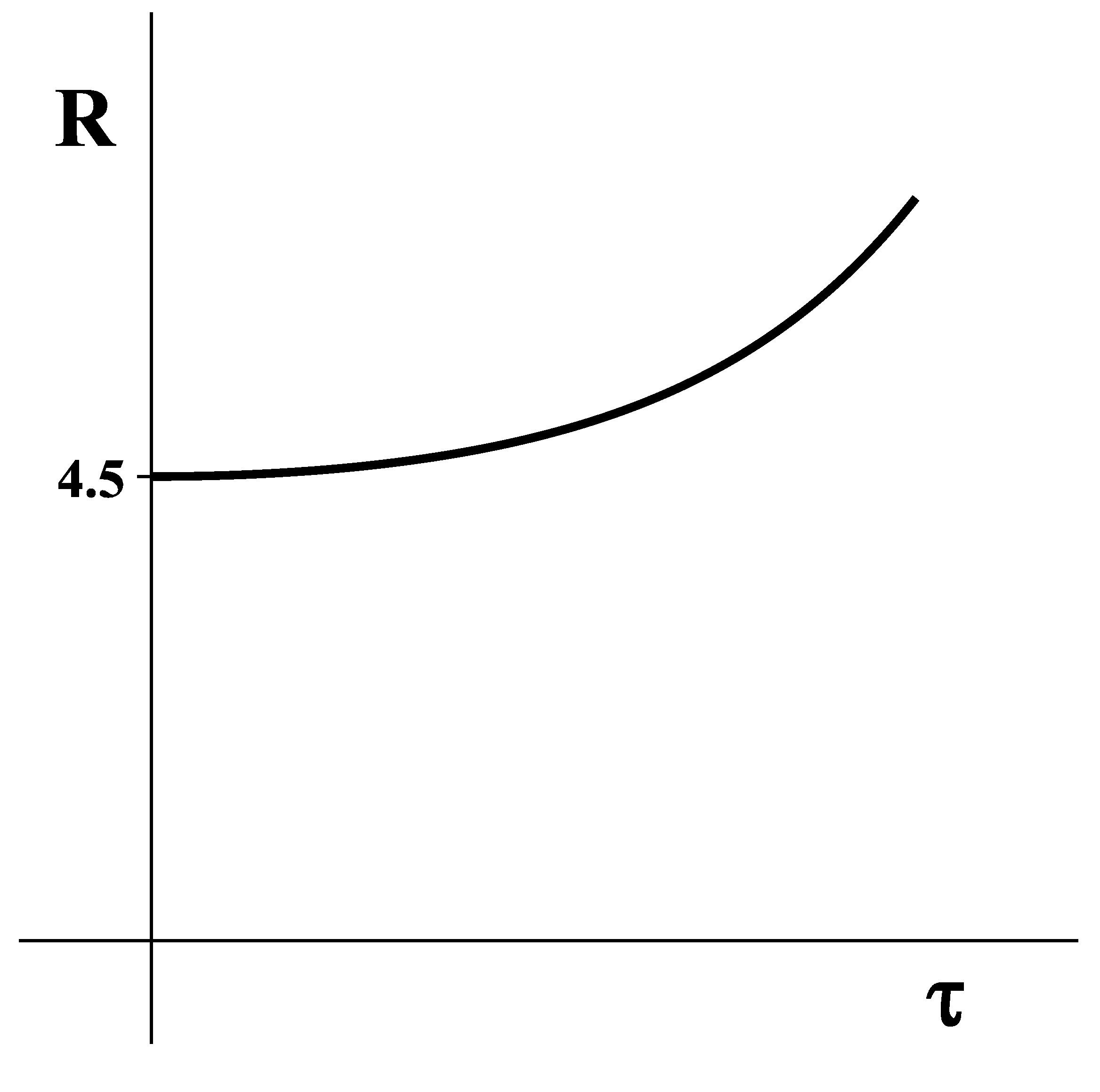}
\caption{\scriptsize  In this figure, the behavior of the radial geodesic of a particle that is dropped at $r=3r_0/2$ is shown; the axes are in units of $r_0$. Notice that, unlike the universe $r\in (\infty,r_m)$, in the universe $r\in (r_m,r_0)$ the particle is pushed away from the throat. }
\label{f24032015a}
\end{minipage}
\end{figure}

Let us study some particular cases. First, consider a radially infalling free particle that comes from infinity ($r\to \infty$) with zero initial velocity. One can easily check from Eq.~(\ref{21032015a}) that these initial conditions imply $E=1$. The behavior of $R$  for this case, Eq. (\ref{20032015d}) with the negative sign (infalling in the region $r>r_m$), is exhibited in Fig.~\ref{f21032015a}. Since $m>-1$, the qualitative behavior of the gravitational field is basically to pull the particle towards $r_0$, which means that when the particle is at  $(r_m,\infty)$ it is attracted by the wormhole throat, but once it is in $(r_0,r_m)$ it never comes back \footnote{Using the plus sign in Eq.~(\ref{20032015d}) one can show that the particle can come back to the universe $r\in (r_m,\infty)$ if an appropriate initial velocity is given to the particle when it is at the other universe.}. It may not be clear in this figure, but there is a subtle change of the concavity of the curve at $(-9.75r_0, 5.84r_0)$, meaning that a repulsive force appears at this point.

To emphasize the fact the particle moves away from $r_m$ when it is in the region $r\in(r_m,r_0)$, the behavior of the geodesic characterized by $\dot{R}=0$ at $r=3r_0/2$ ($R=4.5r_0$) is plotted in Fig.~\ref{f24032015a}. In this case, I have used Eq.~(\ref{20032015c}) to make the plot because  $E=1/\sqrt{3}<1$.

\subsection{The case of $m=-2$ and $n=-1$}\label{s11042015}
For $m=-2$ and $n=-1$, the Eqs.~(\ref{14032015e}) and (\ref{14032015d}) become
\begin{equation}
E^2=A^{-1}+A^{-3}\dot{r}^2, \label{11042015a}
\end{equation}
\begin{equation}
\Delta \tau=\pm r_0\int \frac{dA}{(1-A)^2\sqrt{E^2A^3-A^2}}. \label{11042015b}
\end{equation}
The solution of Eq.~(\ref{11042015b}) is
\begin{eqnarray}
\tau=\frac{\pm r_0}{E^2-1}\Biggl[ 2(E^2-1)\arctan\sqrt{E^2A-1}+\frac{3E^2-2}{\sqrt{E^2-1}}\arctanh \sqrt{\frac{E^2A-1}{E^2-1}}
\nonumber\\
-\frac{\sqrt{E^2A-1}}{A-1} \Biggr]+\textrm{constant}. \label{11042015c}
\end{eqnarray}

From Eq.~(\ref{11042015c}) we get the inequality $A\geq 1/E^2>0$, which means that the particle cannot reach $r_0$ by following a geodesic; it would need an infinite energy to do so. This is probably due to the repulsive nature of the gravitational field. In addition, $E^2$ must be larger than $1$. This follows from the combination of $A\geq 1/E^2$ with $A<1$. As a result, the particle cannot have zero velocity at infinity [notice from Eq.~(\ref{11042015a}) that this would imply $E=1$].

\begin{figure}[h]
\includegraphics[scale=0.25]{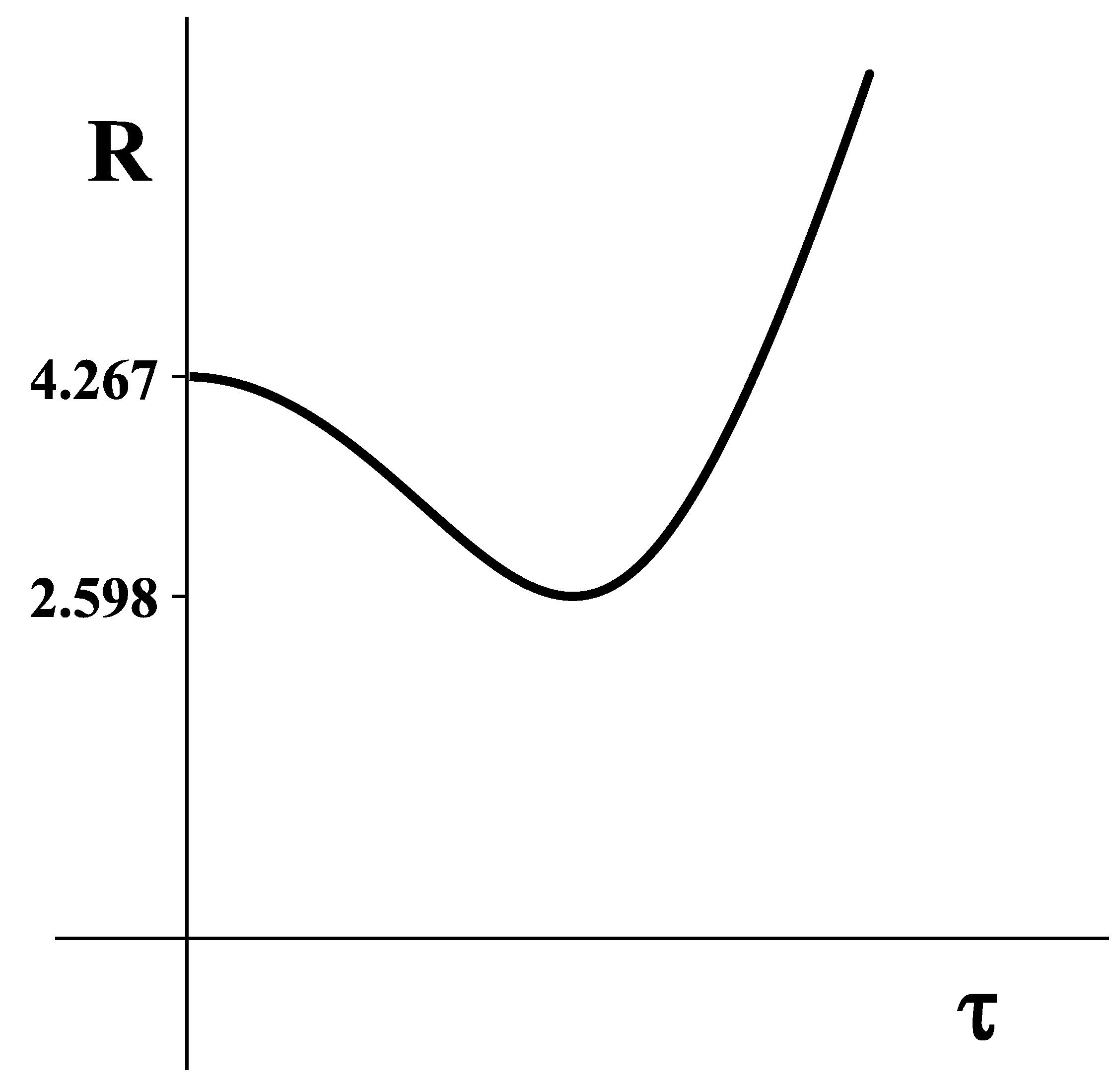}
\caption{\scriptsize  This figure shows the radial geodesic of a particle that is dropped at $r_i\approx 1.07r_0$ ($R\approx 4.267r_0$). For simplicity, I have set $r_0=1$ in the plot. The regions $r_0<r<r_m$ and $r>r_m$ correspond, respectively, to the left and the right sides of the minimum of the curve. As this figure suggests, the particle is attracted to the region $r>r_m$.}
\label{f11042015a}
\end{figure}

Since $n$ is negative, we have a wormhole whose throat is at $r_m=3r_0/2$. The energy needed so that the particle pass through this throat can be obtained from the condition  $A\geq 1/E^2$. This condition is equivalent to 
\begin{equation}
r\geq\frac{E^2}{E^2-1}r_0. \label{18082015a}
\end{equation}
It is obvious that the particle will pass through $r_m$ only if $E^2/(E^2-1)<3/2$  (the minimal value of $r$ has to be smaller than $r_m$), which implies $E^2>3$. It is interesting to note that the smallest value of $r$ is exactly where $\dot{r}=0$ ($\dot{R}=0$). As an example, let us take $E=4$, which is equivalent to taking $\dot{R}=0$  at $r=16r_0/15\approx 1.067r_0$. We already know that for $m<-1$ we must have $r\geq r_i$ (repulsive in the sense of $r$), where $r_i$ is the place where $\dot{r}=0$. Nonetheless, this does not mean that the filed is repulsive in the sense of $R$. As it is clear in Fig.~\ref{f11042015a}, once the particle is dropped at $r=16r_0/15$  ($R=4.267r_0$), it is repelled from $r_0$. Nonetheless, from $r_i$ to $r_m=3r_0/2$ ($R\approx 2.598r_0$) the coordinate $R$ is decreasing, while in the region $r>r_m$ it is increasing in time. In terms of energy, the cases $E^2<3$, $E^2=3$, and $E^2>3$ corresponds to $\dot{R}=0$ at $r>r_m$, $r=r_m$, $r<r_m$, respectively. In another words, the particle is pushed from the universe characterized by $r_0<r<r_m$ to  the other universe, $r>r_m$.

\subsection{The case of $m=-1$ and $n=0$}\label{s30042015}
For $m=-1$ and $n=0$, we have $r=R$ and Eq.~(\ref{14032015e}) takes the form
\begin{equation}
E^2=1+A^{-1}\dot{R}^2, \label{20082015a}
\end{equation}
If we set $\dot{R}_i$ equals to zero at some $R_i$ different from $r_0$, then, from Eq.~(\ref{20082015a}), we see that $\dot{R}$ is constant and the particle must remain at $R_i$. On the other hand, if the particle is moving, it will always reach $r_0$ with zero velocity. Furthermore, from the analysis made in Sec.~\ref{sc}, we can say that the gravitational field will be repulsive. 

\begin{figure}[h]
\includegraphics[scale=0.25]{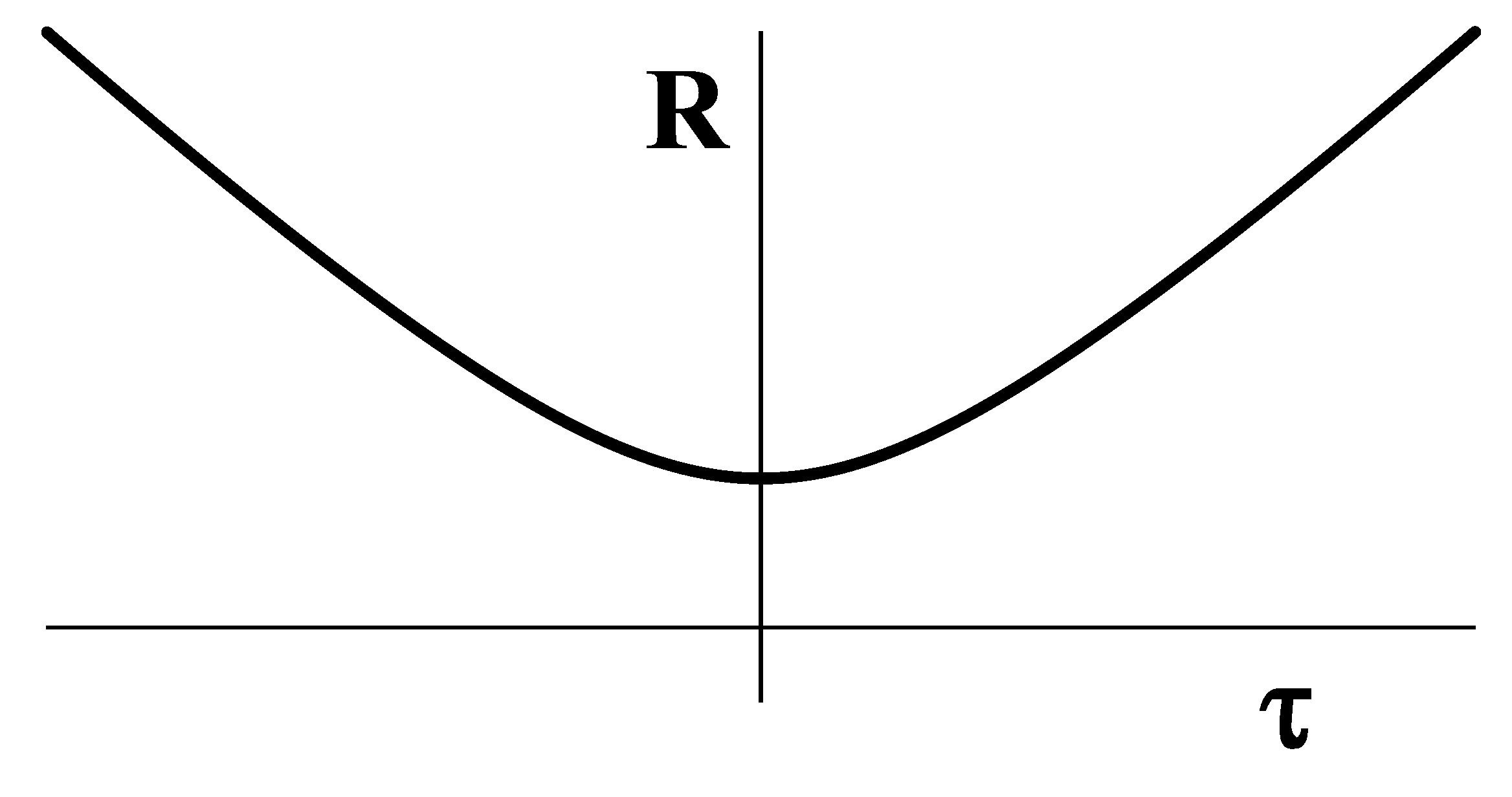}
\caption{\scriptsize  In this plot, the solid curve shows the qualitative  behavior of Eq. (\ref{30042015c}) for the case $E^2=2$. Note that the particle reaches $r_0$ (the minimal value of $R$) at a finite time and with a zero velocity. It is also clear that the gravitational field is repulsive. Since $r_m=r_0$ and $r\geq r_0$, there is no wormhole in this case.}
\label{f20150430a}
\end{figure}

The solution of Eq.~(\ref{20082015a}) for $A>0$ takes the form

\begin{equation}
\tau=\pm \frac{r_0}{\sqrt{E^2-1}}\left[ \frac{\sqrt{A}}{1-A}+\arctanh\sqrt{A} \right]+\textrm{constant} \label{30042015c}
\end{equation}
The behavior of this equation for $E^2=2$ (notice that, for $A>0$, $E>1$) is exhibited in Fig.~\ref{f20150430a}.

In the case treated in this section, there is a change in topology when passing through $r_0$. Therefore, we must take $r\geq r_0$. As $r_m$ is equal to $r_0$, there will be no wormhole.

\section{Final remarks}\label{sd}
In this paper I have studied some properties of the family of solutions (\ref{05032015d}) and found the values of $m$ and $n$ that may allow one to take $r\in (0,r_0)$, although I have analyzed only the cases $r\geq r_0$ (the case $A<0$ will be left for a future paper). I have also found the complete set $\{(m,n)\}$ that avoids singularity in the invariants (\ref{1032015a})-(\ref{1032015c}) at $r_0$, namely, $(m,n)\in \{(0,0),\ (-1,0),\ n\leq -1\}$. 

We have seen that for negative values of $n$, the spacetime (\ref{05032015d}) satisfies the so-called ``strong flare-out condition'' given by Hochberg and Visser. This means that these spacetimes possess wormholes, as it had already been proved in Ref. \cite{PhysRevD.86.084031}.

By studying the behavior of the radial geodesics, I have discovered that in the case $m>-1$ the particle will be attracted to $r_0$, while for $m<-1$ it will be repelled. Examples of each case were given in Secs.~\ref{s22032015a} and \ref{s11042015}.  The case $m=-1$ was also studied and the results showed that the gravitational field repels particles away from $r_0$ for $n\leq 0$, attracts them for $n\geq 1$, and has two possible cases depending on whether $r$ is bigger or smaller than a certain $r_c$ for $0<n<1$: it attracts particles to $r_0$ for $r<r_c$, and repels them for $r>r_c$. It was also possible to analyze the behavior of $R$ as a function of the proper time and identify many properties of the geodesic motion.


%

\end{document}